\documentstyle[12pt,aaspp4]{article}
\begin{document}

\title{Ultraviolet Imaging of the Irregular Galaxy NGC 4449 with UIT: 
Photometry and Recent Star-Formation History}

\author{Robert S. Hill\altaffilmark{1,2},
Michael N. Fanelli\altaffilmark{1,2},
Denise A. Smith\altaffilmark{3},
Ralph C. Bohlin\altaffilmark{3},
Susan G. Neff\altaffilmark{2},
Robert W. O'Connell\altaffilmark{4},
Morton S. Roberts\altaffilmark{5},
Andrew M. Smith\altaffilmark{2},
Theodore P. Stecher\altaffilmark{2}}
\altaffiltext{1}{Raytheon STX Corp., Code 681, NASA/Goddard Space Flight
Center, Greenbelt, MD 20071}
\altaffiltext{2}{Laboratory for Astronomy and Solar Physics, Code 681,
NASA/GSFC, Greenbelt, MD 20771}
\altaffiltext{3}{Space Telescope Science Institute, 3700 San Martin Dr.,
Baltimore, MD 21218}
\altaffiltext{4}{University of Virginia, P.O. Box 3818,
Charlottesville, VA 22903}
\altaffiltext{5}{National Radio Astronomy Observatory, Edgemont Rd.,
Charlottesville, VA 22903}

\begin{abstract}  
The bright Magellanic irregular galaxy NGC 4449 was observed during the
Astro-2 Space Shuttle mission by the Ultraviolet Imaging Telescope (UIT),
which obtained images of a $40\arcmin$ field centered on the galaxy in
two broad far-ultraviolet (FUV) bands centered at 1520~\AA~ and
1620~\AA, with $3\arcsec-5\arcsec$ spatial resolution.  Together with 
H$\alpha$ and H$\beta$ fluxes from ground-based Fabry-Perot images, these
data are analyzed in order to explore the recent star formation history
of NGC 4449.   Maps of the flux ratios H$\alpha$/FUV and FUV/blue
continuum are presented and interpreted using evolutionary synthesis
models.   Photometry is presented both for 22 apertures containing large
OB complexes and for 57 small apertures containing compact FUV-emitting
knots.  The OB complexes along the northern edge of the visible system
have high H$\alpha$/FUV ratios, and thus appear to be more dominated by
the current generation of stars than are other parts of the galaxy.
However, young sources do exist elsewhere and are particularly
conspicuous along the bar.  The small aperture analysis shows three
candidate regions for sequential star formation.  Surface brightness
profiles are consistent with an exponential disk in both the FUV and the
optical continuum.
\end{abstract}


\section{Introduction\label{sec:intro}}

The bright, nearby irregular galaxy NGC 4449 is a superb target for
observation at FUV wavelengths ($\lambda \lesssim 2000$ \AA).  As
observed with broad-band filters, the morphology of NGC 4449 consists
of large OB association complexes against a smooth background.  The
surface brightness peaks at a distinct nucleus.   Balmer emission-line
observations show that the OB complexes are embedded in a welter of
ionized filaments (\cite{sabba2}).  Both the ionized and the neutral
gas show interesting morphology.  The innermost H{\sc i} system,
corresponding to the optically bright part of the galaxy, has roughly
the form of a ring $\sim 5\arcmin$ in diameter (\cite{hg89}).  At
greater radii, NGC 4449 is surrounded by a vast H{\sc i} halo with a
diameter of $\sim 2^\circ$, including a number of long, almost
straight, armlike filaments (\cite{huntwilc}).  Because of such traits,
NGC 4449 has been studied frequently in recent years.

The type given for NGC 4449 in the {\em Third Reference Catalogue of
Bright Galaxies} (RC3; \cite{rc3}) is IBm (barred Magellanic
irregular), and that given in the {\em Revised Shapley-Ames Catalog}
(RSA; \cite{rsa}) is SmIV (low-luminosity, very late spiral).  The
distance adopted for this paper is 5.4 Mpc (\cite{kraan}).  NGC 4449 is
well-suited observationally for the investigation of stellar and
interstellar structures on scales of $10^{2}$ to a few $10^{3}$ pc,
which correspond to angular distances of a few arcsec up to several
arcmin.

An earlier paper (\cite{rocket}; Paper I) described an FUV image of NGC
4449 made with a sounding-rocket telescope (flight 36.068GG from White
Sands) and compared it to optical images taken in narrow bands with the
Goddard Astronomical Fabry-Perot Imaging Camera (GAFPIC; formerly known
as GFPI).  The GAFPIC data included both H$\alpha$ and H$\beta$, as
well as nearby continuum bands.  Paper I discussed in detail the
relationship between continuum and emission-line images, both in a
qualitative, morphological sense, and quantitatively.  In particular,
the ratio of H$\alpha$ to FUV emission in 22 OB association complexes
was used to investigate recent star formation history, by means of
association ages computed from evolutionary synthesis models of young
stellar populations.  However, the spatial resolution and photometric
precision of the FUV data were degraded somewhat by a detector
malfunction.  Both for this reason and because of the short duration of
the suborbital observation (126 s), it was decided to reobserve NGC
4449 using UIT.

The UIT observation was carried out in March, 1995, during the Astro-2
mission of the Space Shuttle {\em Endeavour} (STS-67).  The deepest UIT
exposure (B1 filter, 987 s) reaches a limiting FUV surface brightness
fainter than $\sim 2\times 10^{-18}$ ergs cm$^{-2}$ s$^{-1}$ \AA$^{-1}$
arcsec$^{-2}$, which is $\lesssim 20\%$ of that reached by the sounding
rocket image.  The spatial resolution of this image is $\sim 5\arcsec$,
comparable to the rocket image.  However, another UIT image (B5 filter,
493.5 s) has a better resolution of $3\arcsec$ because of the shorter
exposure, since slight trailing occurred during the observation
sequence.  

All of the UIT images have a smoother, more sharply peaked, and better
determined PSF than the rocket image, because the rocket data, taken
with a multi-anode microchannel array (MAMA) detector, required image
reconstruction from time-tagged photon events to compensate for the
combined effects of severe trailing and of electronics damage at
launch.  For the same reason, the flat fielding of the reduced UIT
images is superior.  Also, the UIT observational program included a
planned suite of calibration observations, whereas the rocket
calibration was based on archival {\em IUE} spectra of objects in the
science image; hence, the calibration of the UIT images is better
determined.

Although we have new FUV data, the optical data used in this paper are
the same as those reported in paper I.  However, comparison to other
H$\alpha$ photometry (\cite{mayya}) resulted in finding a mistake in
the calibration used in Paper I.  Therefore, we present new photometry
in optical bands as well as FUV, with recomputed  ages for the OB
complexes.  In addition, we extend the analysis given in Paper I.
First, we present color maps showing the spatial distributions of two
flux ratios: H$\alpha$/FUV, and FUV/optical continuum.  Second, we
analyze the H$\alpha$/FUV ratios of 57 FUV-bright knots $\sim 100$ pc
in diameter, smaller than the large OB complexes measured in Paper I. 
Integrated photometry and surface photometry are also presented.  

The contents of this paper are as follows: 
\S\ref{sec:intro},introduction; 
\S\ref{sec:obs}, observations; 
\S\ref{sec:glob}, morphology, integrated photometry, flux ratio
maps, and surface photometry;
\S\ref{sec:sfr}, aperture photometry and star formation history;
\S\ref{sec:conc}, conclusions.

\section{Observations\label{sec:obs}}

\subsection{Overview\label{sec:obs:ovv}}

UIT is a 38 cm Ritchey-Chr\'{e}tien reflector with two separate
UV-sensitive detectors, selectable by a rotating pick-off mirror. Each
detector is an image tube coupled to Eastman Kodak II-aO film by fiber
optics; the FUV camera has a CsI photocathode, and the near ultraviolet
(NUV) camera has a Cs$_{2}$Te photocathode. The observations discussed
in this paper are in two of the FUV bands, i.e., B1 (CaF$_{2}$ filter,
$\lambda =1520$ \AA, $\Delta \lambda =350$ \AA) and B5 (SrF$_{2}$
filter, $\lambda =1620$ \AA, $\Delta \lambda =220$ \AA).  The B1 and B5
bandpasses differ mainly at the short-wavelength cutoff, which is 1300
\AA~ for B1, and 1460 \AA~ for B5. B1 has the higher throughput, but it
is useful only during orbital night, because the bandpass includes
strong O{\sc i} airglow lines (e.g., \cite{airglow}; \cite{airwalr}). 
B5 is useful both for orbital daytime and for emphasizing C{\sc iv}
emission.

UIT Astro-2 data reduction includes digitization with a
microdensitometer, adjustment of film fog level, linearization with a
characteristic curve, and flat fielding. Astrometry is done for fields
where enough FUV-emitting field stars are available. This paper uses
the final version of the UIT Astro-2 data reduction pipeline, which is
called by the name FLIGHT22.  Detailed descriptions of the UIT
instrument and data reduction are given elsewhere (\cite{uit};
\cite{stecher92}). The Astro-2 UIT calibration is based on 75
measurements of 49 stars that have been observed by {\em IUE}. UIT
fluxes are tied to the Finley {\em IUE } calibration (\cite{iuecal}),
which is similar to the  calibration for the final {\em IUE} archive,
known as NEWSIPS (\cite{newsips}). Error in tying UIT to {\em IUE} is
$\sim 10\%$ (\cite{calerr}).

Our observations of NGC 4449 are listed in Table \ref{tab:obs}. The UIT
images were obtained using the FUV camera during the Astro-2 mission;
the NUV camera failed at launch. In this paper, the $986$ s B1 image
and the $493.5$ s B5 image are discussed.  Before analysis, a sky value
of $9.13 \times 10^{-20}$ ergs s$^{-1}$ cm$^{-2}$ \AA$^{-1}$
arcsec$^{-2}$ is subtracted from the B1 image (orbital night) and $2.60
\times 10^{-18}$ ergs s$^{-1}$ cm$^{-2}$ \AA$^{-1}$ arcsec$^{-2}$ from
the B5 image (orbital day).  Surface brightness can be expressed in
monochromatic magnitudes using the formula $\mu = -2.5 \log
(f_\lambda/A) -21.1$, where $f_\lambda$ is the mean flux per pixel and
$A$ is the area of a pixel in arcsec$^2$.  In these units, the B1 sky
is 26.50 mag arcsec$^{-2}$, and the B5 sky is 22.86 mag arcsec$^{-2}$.

NGC 4449 was observed from the ground using GAFPIC (\cite{gafpic}).
These observations, made by K.-P. Cheng, R. Oliversen, and P. M. N.
Hintzen, were described in detail in Paper I. The GAFPIC data consist of
Fabry-Perot emission-line images at H$\alpha$, H$\beta $, and nearby
continuum wavelengths with bandwidths of $20-25$ \AA.  Hereafter, the
continuum in a narrow band near H$\beta$ is called {\em blue}, and near
H$\alpha$, {\em red}.  {\em H$\alpha$} and {\em H$\beta$} always refer
to the continuum-subtracted data.

\subsection{Calibration Changes in Earlier Data\label{sec:obs:cal}}

A systematic error in the rocket FUV fluxes in Paper I was discovered
through a comparison with UIT data.  Although the cause of the
discrepancy was not determined, one possibility was a mistake in the
coordinates logged in the {\em IUE} observing records for the archival
spectra used for calibration.  These spectra, SWP 5663 and LWP 4908,
were supposed to be of the nucleus of NGC 4449; however, we conjecture
that the actual target may have been another bright knot $0\farcm5$ to
the south.  In any case, the effect of the error was to make the FUV
fluxes too high by a factor of $\sim 2$.  The corrected rocket
calibration is consistent with that for an earlier flight using the
same detector (\cite{ngc6240}).

Many of the arguments of Paper I relied on the H$\alpha$/FUV flux ratio.
Coincidentally, the FUV calibration error was nearly canceled in this
ratio by a similar error in the H$\alpha$.  The plausibility of the
combined results hindered the discovery of both problems.  In fact, all
of the optical fluxes reported in Paper I were a factor of $1.7$ too
high, because an incorrect aperture diameter was applied to the H{\sc
ii} region HK11 (\cite{kenncal}), whose H$\alpha$ flux was the
calibration standard.

The corrected calibration of the GAFPIC data has been confirmed by
comparison to another study that also uses HK11 as the calibration
source (\cite{mayya}).  Also, the GAFPIC images have been refined by
subtracting ghost images reflected from the Fabry-Perot blocking
filter.  The effect of the ghosts on the photometry of bright sources
is negligible, and the main result of making the correction is to remove
faint false sources from the sky $1\arcmin - 2\arcmin$ west of the
galaxy bar.  Other aspects of the GAFPIC data reduction that were
described in Paper I are not detailed again here, e.g., the correction
for N{\sc ii} contamination and for the radially varying central
wavelength of the Fabry-Perot sensitivity curve.  However, we note that
the continuum to be subtracted from the line$+$continuum images is
scaled using foreground stars, and that H$\beta$ is calibrated relative
to H$\alpha$ using the H{\sc ii} region spectroscopy of \cite{lequeux}.

The combined calibration errors resulted in values of the H$\alpha$/FUV
flux ratio for individual OB complexes in Paper I that were too small
by $\sim 20\%$.  The significance of this ratio was that in
combination with evolutionary synthesis models, it gave a so-called
instanteous-burst (IB) age for each source.  The systematic error in
this age resulting from the calibration was $\sim 0.5$ Myr.  Since the
range of ages found was $0-10$ Myr, a 0.5 Myr shift does not invalidate
the arguments presented in Paper I on relative ages and  morphology. 
However, the calibration does affect the global star formation rate
(SFR).  Moreover, because the UIT data have better flat fielding and
S/N than the rocket data, we have used the UIT data to produce a
complete revision of the photometric results for the  22 OB complexes
(\S\ref{sec:sfr:lgap:uit}).  The interpretation in terms of star
formation history, while consistent with that in Paper I, is restated
here in a more precise way, using four bandpasses.

\subsection{Extinction \label{sec:obs:ebv}}

To correct for foreground extinction, we use the functional form of the
extinction curve given by \cite{ccm} with the parameter $R_V$ set to
3.1; hereafter, the term {\em Galactic} in reference to extinction
always means this curve.  H$\alpha$/H$\beta$ ratios are converted to
reddenings using the formula
$E(B-V)=-1.00+2.184\log(\mathrm{H}\alpha/\mathrm{H}\beta)$, which is
computed from the Galactic curve.

To correct for extinction within NGC 4449, we use the algorithm of
\cite{calz4}.  This procedure is based on studies of extinction in
starburst galaxies, in which the H$\alpha/$H$\beta$ ratio of the
excited gas is compared to the ultraviolet spectral slope of the
associated stellar component (\cite{calz1}; \cite{calz2};
\cite{calz3}).

The Calzetti algorithm begins with the computation of a Balmer
decrement for each source.  The reddening $E(B-V)_g$ of the ionized gas
is computed directly from the Balmer decrement using the Galactic
extinction curve.  $E(B-V)_g$ is multiplied by $0.44$ to obtain
$E(B-V)_s$, which is the reddening of the associated stars.  Broad-band
continuum fluxes attributable to the stellar component are corrected in
the FUV using $E(B-V)_s$ with a ``starburst'' extinction curve, for
which \cite{calz4} gives a functional form.  This extinction curve is
similar to ones found for the LMC, except that it lacks a 2200\AA\
bump.  Optical emission-line fluxes arising from the gas are corrected
using $E(B-V)_g$ with the Galactic extinction curve.

In most of this paper, the single value $E(B-V)_g=0.43$, including both
foreground and internal extinction, is adopted for all of NGC 4449.
This value is the mean of Balmer-decrement reddenings computed for the
OB complexes analyzed both in Paper I and in this paper below (\S
\ref{sec:sfr:lgap}).  In order to avoid the influence of any error in
sky or continuum subtraction in the low-surface-brightness portions of
the H$\beta$ image, the Balmer decrements are computed only from the
bright pixels.  For each aperture separately, the histogram of
H$\alpha$ fluxes of the pixels is computed.   The 95th percentile is
found, and the pixels brighter than that value are used in both
H$\alpha$ and H$\beta$ to compute the Balmer decrement (Paper I).  A
set of 22 reddenings results, one for each large aperture, and the mean
of this set is adopted as the value for NGC 4449.  The scatter among
the 22 values is  $\sigma=0.1$ mag.  

The Galactic foreground reddening is estimated to be $E(B-V)=0.03$,
computed by applying the dust-to-gas ratio of \cite{dustgas} to the
H{\sc i} column density near the line of sight from \cite{bellsur}.
Therefore, $E(B-V)_g$ in NGC 4449 is 0.40 and $E(B-V)_s$ is 0.18.

The mean Balmer decrement reddening $E(B-V)=0.43$ can be compared with
values from other authors.  \cite{cmartin} tabulates several
logarithmic extinctions at H$\beta$ in NGC 4449.  Converted to
$E(B-V)$, these values are 0.07, 0.5, and 0.2 for diffuse ionized gas
and 0.03 and 1.2 for H{\sc ii} regions.  Spectra in an effective
aperture of $3\arcmin \times 2\arcmin$ were obtained by \cite{kenn92}
using a drift scan.  From the resulting Balmer decrement, we compute
$E(B-V)=0.47$.  In summary, our value appears to be in the appropriate
range.

Our justification for using the Calzetti algorithm is that in
conjunction with evolutionary synthesis models, it yields fairly
consistent SFRs over several timescales, each of which is based on a
different bandpass.  This analysis and the influence of the extinction
law are described below (\S \ref{sec:sfr:lgap:uit}).

\section{Global Properties of NGC 4449\label{sec:glob}}

\subsection{Morphology and Integrated Photometry\label{sec:glob:mph}}

NGC 4449 is surrounded by a neutral hydrogen halo extending to as far
as $\sim 1^\circ$ from the visible system, reaching in projection the
neighbor galaxy DDO 125, as shown by an Effelsberg 100-meter map
(\cite{hihalo}).  The corresponding VLA map (\cite{huntwilc}) shows
that this gas is highly structured, with linear streamers and tails
$\sim 30\arcsec$ long.  The UIT images show no UV-emitting counterpart
to the H{\sc i} halo, so that within a $20\arcmin$ radius of the galaxy
center, star formation appears to be confined to the well-known
features of the visible galaxy.

Subimages of the UIT frames of NGC 4449 are compared with optical data
in Figure \ref{fig:all}.  Following \cite{hk83}, the origin of the
coordinate system is one of the four foreground stars seen in the
optical images.  The UIT images are on the left, and the blue and
H$\alpha$ images are on the right.  Both the FUV and the optical bands
show bright knot complexes against a relatively faint, but still
evident, background.  However, the morphology of the background depends
on whether it is observed in broad-band continuum or in line emission. 
In the H$\alpha$ image, the background is intricately structured and
comprises numerous ionized filaments, but in the FUV and blue continuum
images, the background is relatively smooth and diffuse.  Figure
\ref{fig:cont} shows a contour map of the B1 image. The lowest contour
plotted, which is that for a surface brightness $1.3 \times 10^{-17}$
ergs cm$^{-2}$ s$^{-1}$ \AA$^{-1}$ arcsec$^{-2}$, encloses almost all
of the galaxy, including the diffuse component.

The FUV-emitting knot seen in the western part of the UIT subimages in
Figure \ref{fig:all} lies outside the field of view of the optical
data. The region between this isolated knot and the main body of the
galaxy contains X-ray emitting gas (\cite{xray}).  A ring or torus of
neutral hydrogen is coincident with the galaxy in projection
(\cite{hg89}).  The eastern side of the ring, which has a broken,
somewhat chaotic morphology in H{\sc i}, coincides with the main body
of the galaxy, and the western side of the ring coincides with the
isolated FUV knot.

Conspicuous in NGC 4449 at all of our observed wavelengths are the 
sites of recent star formation, which are found in the apparent bar of
NGC 4449, as well as in the armlike extensions to the north and south. 
Note that the bar of NGC 4449 is different from those found in barred
spiral galaxies, as shown by the $K$-band surface brightness profile
(\cite{thronson}).

The FUV and $B$ absolute magnitudes of NGC 4449 are compared with those
of several other galaxies in Table \ref{tab:integ}.  The FUV magnitudes
are preliminary measurements from the UIT Galaxy Atlas (Fanelli et al.
1998, in preparation), and the $B$ magnitudes are from the RSA.  NGC
4449 is bright for a Magellanic irregular, though not extreme 
(\cite{huntwilc}), and it is of comparable brightness to some 
spirals.  In broad-band continuum magnitudes, both FUV and optical, NGC
4449 is intrinsically $2-3$ times as bright as the LMC.  

The intrinsic FUV$-B$ color of NGC 4449 implied by Table
\ref{tab:integ} is $-1.43$.  This value can be compared to the UV$-B$
colors of RC3 galaxies observed by the FAUST ultraviolet telescope
(Figure 2 of \cite{faust}), where UV$\approx1600$\AA.  The result is
that NGC 4449 is near the mean color for FAUST irregulars.

An important indicator of an unusual history for NGC 4449 is the gas
kinematics, which several studies show to be peculiar, both in the
central, bright system and in the extended H{\sc i} envelope.  By
peculiar, we mean primarily that the observed motions are clearly
inconsistent with a single rotating disk.  The velocity gradient along
the major axis in the inner, optically bright part is in the opposite
sense to the rotation of the outer H{\sc i} system (\cite{sabba2};
\cite{eliot}).  On a local scale, sharp velocity gradients are seen
within the bright ionized gas, and the resulting shocks are a possible
stimulus for star formation (\cite{hgh}).

\subsection{Flux Ratio Maps\label{sec:glob:rat}}

\subsubsection{General Interpretation\label{sec:glob:rat:gen}}

Flux ratio maps have considerable heuristic value in the study of
galaxies.  Continuum ratios can show spatial patterns both of star
formation and of dust, because the slope of the aggregate spectrum is
reddened either by extinction  or by the aging of the stellar
population.  Ratios involving emission lines can show how the
excitation processes in the interstellar medium are distributed in
relation to stars.  Here, we investigate NGC 4449 using two ratio maps,
H$\alpha$/B1 and B1/blue continuum.

An important tool for analyzing flux ratio maps, as well as the
aperture photometry described later in this paper, is evolutionary
synthesis modeling.  Fluxes from a cluster (or any coeval group) of
stars are computed by a weighted sum of contributions from various
model atmospheres.  The weighting is derived from an assumed initial
mass function (IMF); in this paper, we use a Salpeter power law slope,
which may be expressed as $\gamma=-2.35$, $\Gamma=-1.35$, or $x=2.35$,
depending on the formalism adopted.  The effective temperature and
bolometric flux of a star with a given initial mass are determined as a
function of elapsed time since the formation of the group using
evolutionary tracks.  We use the Geneva tracks (\cite{tracks}) for
$\log Z/Z_\odot=-0.4$ and the stellar atmospheres of \cite{kurucz} for
$\log Z/Z_\odot=-0.3$.  To combine these inputs into evolving cluster
models, we use code written in the IDL language by W. Landsman (Paper
I).  Most results in this paper are based on IB models, which assume a
single delta-function star formation episode.  However, the integrated
galaxy photometry (\S\ref{sec:sfr:lgap:uit}) is analyzed using
continuous star formation (CSF) models, which assume that stars have
been forming at a constant rate since the galaxy came into being.

\subsubsection{H$\alpha$/B1\label{sec:glob:rat:hab}}

Although FUV and H$\alpha$ images both show the locations of early-type
stars, they do not show exactly the same stars.  In a zero-age
population, H$\alpha$ is emitted by interstellar gas photoionized by
early- to mid-O stars, whereas FUV is emitted by stars over a somewhat
broader range of spectral types from O to early B. The masses of the
stars that dominate ionizing flux, $N_{{\rm Lyc}}$, and the 1500 \AA\
continuum, $L_{150}$, have been computed for a Salpeter IMF extending
up to 110 $M_\odot$ by means of the evolutionary synthesis models, with
the following results:  at zero-age, 90\% of $N_{{\rm Lyc}}$ is
contributed by stars with $M/M_\odot > 30$ and lifetime less than $\sim
7$ Myr, and  90\% of $L_{150}$ is contributed by stars with $M/M_\odot
> 12$ and lifetime less than $\sim 30$ Myr.  Hence, OB associations
from $7$ to $30$ Myr old should be much brighter in FUV than in
H$\alpha$ images, provided that they were formed in single bursts of
star formation and no current generation of high-mass stars exists to
contribute ionizing flux.

Even for OB associations less than $7$ Myr old, $N_{{\rm Lyc}}$ and
$L_{150}$ decline at different rates, so that the ratio of H$\alpha$ to
FUV surface brightness can be used as a clock. Unfortunately,
evolutionary synthesis models cannot give this flux ratio directly,
because the configuration of the gas and dust that surround each OB
association is typically unknown.  However, simplifying assumptions can
be made.  The models used in this paper incorporate the following
assumptions:  (1)  H{\sc ii} regions are ionization bounded;  (2)
extinction of the ionizing continuum by dust within H{\sc ii} regions
is negligible;  (3) extinction outside the H{\sc ii} regions is
described by the starburst reddening model of Calzetti (1997) (\S
\ref{sec:obs:ebv}).

Of these three assumptions, the first is known to be inaccurate. 
Leakage of ionizing photons from NGC 4449 H{\sc ii} is estimated at
$\sim 20\%$ (\cite{hg92}); moreover, in a study of the transition from
H{\sc ii} regions to diffuse ionized gas, \cite{cmartin} finds no clear
boundary between these two domains.  If the leakage is consistent over
all of the H{\sc ii} regions, the sequence of relative ages computed
from H$\alpha$/FUV ratios will not be significantly affected.  The
zero-point shift resulting from the detection of only 80\% of ionizing
flux is $\sim 0.5$ Myr.  The second assumption, that of low internal
extinction in the H{\sc ii} regions, is consistent with the Calzetti
method of extinction correction, which implies that the extinction is
caused primarily by dust in a shell configuration, i.e., dust that lies
in front of the sources rather than being mixed with them
(\cite{calz3}).

Figure \ref{fig:model} plots the logarithm of the theoretical $N_{{\rm
Lyc}}$/B1 ratio vs. age, computed from the IB models as described
above.   The Y axis on the right-hand side shows the observable
quantity H$\alpha$/B1 that results if $E(B-V)=0.43$ is applied as
described above (\S \ref{sec:obs:ebv}).  This plot can be compared with
the grayscale map in Figure \ref{fig:rathau}, which shows the spatial
distribution of the observed $H\alpha$/B1 ratio.  Before taking the
quotient, the H$\alpha$ image is smoothed with a Gaussian of FWHM
$3\farcs 6$ to match the B1 resolution.

The black areas in Figure \ref{fig:rathau} correspond to IB ages of
$\lesssim 4$ Myr.  We emphasize that the delta-function star formation
history is unlikely to be a good approximation for every source,
particularly in interpreting the optical continuum, but it does give a
convenient measure of domination by short-lived stars as compared to
stars with lifetimes of a few $10^7$ yr.  The northeast periphery of
NGC 4449, at the upper left of the image, is clearly the most active
region of the galaxy in terms of the dominance of current over past
star formation.  In the rest of the galaxy, smaller star-forming
regions are interspersed with areas of relative quiescence, which are
light in the image.  The two most conspicuous of these areas are at
coordinates $(0.7, 0.9)$ and $(1.0, -0.6)$.  The first area appears in
Figure \ref{fig:rathau} as a white crescent in the upper-right quadrant
of the galaxy, and it coincides exactly with an OB association seen in
Figure \ref{fig:all}.  This source is an example of an association
whose age is $\gtrsim 7$ Myr, with values of $\log$ H$\alpha$/B1
$\approx 0.4-0.45$.  The second area appears in Figure \ref{fig:rathau}
as a compact white spot in the southern bar of the galaxy.  Here, the
lowest-valued pixels in $\log$ H$\alpha$/B1 are $\sim 0.25$,
corresponding to an IB age of $\sim 11$ Myr.  Immediately to the west
is a conspicuous knot of current star formation.  It is tempting to
speculate that sequential star formation (SSF) has occurred in this
region, proceeding from east to west (\S\ref{sec:sfr:smap}).

A possible alternative explanation for the high H$\alpha$/B1 ratios is 
extinction, which would preferentially suppress the FUV.  A difference
in Balmer decrement reddening of $\Delta E(B-V) = +0.1$ increases
$\log$ H$\alpha$/B1 by $0.1-0.2$, depending on the extinction model,
with the Calzetti starburst model at the low end of this range and the
Galactic extinction curve at the high end. Thus, variations of this
size could result from neglected variation in $E(B-V)$, since we use an
average value.  However, the range of values seen in Figure
\ref{fig:rathau} is much greater, $\pm 0.8$ in $\log$ H$\alpha$/B1. 
This question is considered in greater detail below (\S
\ref{sec:sfr:lgap:uit}).

\subsubsection{B1/Blue\label{sec:glob:rat:b1b}}

A plot of the B1/blue ratio from an aging OB association is shown in
Figure \ref{fig:long}.  This flux ratio is computed from IB models as
described above; for comparison, an H$\alpha$/B1 model is also shown.
The left-hand Y axis gives unreddened ratios, and the axes on the right
give the ratios as reddened by our average extinction model
(\S\ref{sec:obs:ebv}).  The corresponding map of the B1/blue ratio
observed in NGC 4449 is shown in Figure \ref{fig:ratuvb}.  To match
resolutions, the  blue is smoothed with a Gaussian of FWHM $3\farcs 6$,
then a residual difference in PSF shape is removed by smoothing both
images with a Gaussian of FWHM $2\farcs3$.   Dark areas have a high
B1/blue ratio and therefore represent stars relatively young on the
average.  The spatial variations are fairly smooth, and the map is
dominated by broad features of order $\sim 0\farcm5$ in diameter, a few
times the smoothed resolution.

Of interest in Figure \ref{fig:ratuvb} is the absence of any strong
features corresponding either to the nucleus of NGC 4449 (coordinates
$0.8, -0.1$; cf. Figure \ref{fig:all}) or to two FUV sources that lie
$\sim 0.3$ arcmin to the northeast of the nucleus.  The H$\alpha$ image
and the H$\alpha$/B1 ratio image show that star formation is happening
currently in these knots.  However, they are in the bar, which has the
highest broad-band surface brightness at all wavelengths.  Most likely,
this region is the site of frequent star formation episodes and has a
historically high mean star formation rate, as compared to the
periphery.  As a result, relatively recent star formation in the bar
fails to outweigh older generations of stars in the continuum flux
ratio map, whereas at larger radii, star formation episodes are
infrequent, and one recent generation can outshine the older stars left
from previous episodes.

\cite{hg90} show that the northern periphery of NGC 4449 coincides with
an H{\sc i} cloud, which may feed the star formation.

\subsection{Radial Profiles\label{sec:glob:rad}}

The radial surface brightness profile of NGC 4449  is shown in Figure
\ref{fig:sbprof} in both the UIT B1 band and the narrow red continuum
band.  The red profile appears to be exponential.  The B1 profile is
lumpy, because of the bright star-forming complexes, which reflect
local conditions in the interstellar medium; however, it appears
approximately exponential except in the center, where a cusp is seen in
both the optical and the FUV.

A similar analysis has been done with UIT data for the irregular galaxy
NGC 4214 (\cite{ngc4214}). These authors find that NGC 4214, otherwise
similar to NGC 4449, shows a deVaucouleurs profile in the FUV out to a
radius of $\sim 3$ kpc.  They argue that because this profile is
characteristic of dynamically hot systems, a merger may have played an
important role in causing the recent star formation in the central
region of NGC 4214.  Although a merger scenario has been suggested in
order to explain the peculiar gas kinematics in {\em both} NGC 4449 and
NGC 4214 (\cite{hgh}), NGC 4449 outside the nucleus appears to be
dominated by a star-forming disk.

\subsection{Froth\label{sec:glob:fro}}

The term {\em froth\/} has been coined to describe a certain morphology
of H$\alpha$-emitting material outside typical H{\sc ii} regions. Froth
lacks conspicuous embedded OB complexes, and it is distinguished from
simple diffuse emission by a complex structure of bright filaments. The
froth is thought to be the product of a mixed ionization mechanism,
with energy supplied mostly by ionizing stars, but also by shocks
(\cite{hg90}; \cite{hg97}).

A plausibility argument for the importance of photoionization can be
made with a simple application of the FUV image data. We examine the
ratio between H$\alpha$ surface brightness in any given local area and
the coincident FUV surface brightness. If the ratio of H$\alpha$ to FUV
is consistent with that expected from an ionizing cluster, then
photoionization is a plausible mechanism. Apertures $\sim 20\arcsec$ in
radius are superposed on 4 regions of diffuse, filamentary H$\alpha$
emission and extended FUV emission (Figure \ref{fig:froth}).  The
H$\alpha$ and FUV fluxes for these regions are given in Table
\ref{tab:froth}, together with $\log N_{{\rm Lyc}}/L_{{\rm B1}}$,
without extinction corrections or background flux subtraction. These
values all fall within a range consistent with IB cluster models, i.e.,
consistent with photoionization by young stars. The FUV surface
brightnesses are equivalent to $\sim 1$ late B star per pc$^{2}$,
as computed using a standard UV stellar library (\cite{fanelli}).
However, this number represents a lower limit, because of extinction. 
An argument based on H$\alpha$ emission measures gives an estimate of
ionizing flux equivalent to $1$ early B star per pc$^{2}$ in some
of the H$\alpha$ structures in NGC 4449 (\cite{hg90}).

More recently, a detailed spectroscopic analysis (\cite{hg97}) has
shown that H$\alpha$ froth structures on all size scales are dominated
by photoionization.  The largest filaments are widely separated from
the main stellar component of NGC 4449, so that their photoionization
requires Lyc photons to traverse kpc-scale paths from the nearest OB
associations.  However, the four regions measured here are embedded in
the luminous body of the galaxy, so that it is not clear whether the
ionization is caused by the conspicuous OB associations or by dispersed
B stars.

\section{Recent Star-Formation History\label{sec:sfr}}

\subsection{Large Apertures\label{sec:sfr:lgap}}

\subsubsection{Recapitulation of Previous Analysis \label{sec:sfr:lgap:rec}}

In Paper I, an FUV sounding-rocket image was used together with the
GAFPIC emission-line imagery to compute ages for emission knots in 22
large apertures of irregular shape with diameters $\sim
0\farcm2-0\farcm5$, each enclosing a complex of OB associations and
H{\sc ii} regions.  After correcting for extinction and converting
H$\alpha$ to $N_{{\rm Lyc}}$ using case B recombination (\cite{agn2}),
the log of the ratio $N_{{\rm Lyc}}/L_{{\rm FUV}}$ provided a measure
of the age via an IB model of an evolving cluster.   It was found that
individually, both the numerator and denominator declined rapidly after
zero-age, so that the ratio was dominated by the youngest stars
included.  Therefore, the modeled IB age gave, to some degree of
approximation, the age of the most recent generation of stars within
each photometric aperture.

Photometry of the large apertures showed high H$\alpha$/FUV ratios for
a group of sources arranged in an arrowhead or bow-wake shape along the
northern periphery of NGC 4449, coinciding with an H{\sc i} cloud.  The
IB model showed that the age difference from the OB complexes in the
bar was $\sim 2.5$ Myr.  The question was whether this was a
significant result.  In other words, could one conclude that the
northern part of NGC 4449 was undergoing a coherent star formation
event, separated in time from the most recent star formation in the
bar?

In order to answer this question, Paper I investigated whether
differences in star formation history could perturb the IB age so that
it failed to give the age of the latest generation of stars, and
instead was contaminated by previous generations.  The conclusion was
that this effect could occur.  An IB model with age $t$ Myr gave the
same value of $\log N_{{\rm Lyc}}/L_{{\rm FUV}}$ as a model of CSF
beginning at the formation of the galaxy and stopping $t-2$ Myr ago. 
The application to the age difference between the north and the bar was
this:  if we applied the IB model to the north and the CSF model to the
bar, then the age difference became $\sim 0.5$ Myr for the youngest
generation, which was not significant.  The age difference was
preserved only if we could assume the same type of star formation
history for both regions.  However, the difference in morphology and
surface brightness between the north and the bar made it seem possible
that two different histories did apply.

Therefore, the interpretation of Paper I ended in a disjunction. 
Either of two interesting results could be true, but the data and
simple models could not decide between them.  Either the north and the
bar had the same type of star formation history, and the latest
generation of stars was younger in the north; or else the latest
generation of stars was the same age in both regions, but they had
different prior star formation histories.  Paper I chose the second
interpretation as more likely.  In either case, the sources measured in
the north differed from those in the bar in a way that was manifested
in the observed H$\alpha$/FUV ratio.

\subsubsection{New Analysis Including UIT Data\label{sec:sfr:lgap:uit}}

Table \ref{tab:lgpos} gives the positions and sizes of the 22 large
apertures defined in Paper I.  These apertures are used for  photometry
of both the UIT B1 data and the recalibrated GAFPIC data. The large
apertures are shown in Figure \ref{fig:lgap}; each encloses one major
complex of OB associations.  The main result of Paper I was a tendency
for the knots in the bar region to have greater IB ages than those in
the outlying regions along the northern edge of the galaxy by $2-3$
Myr.  Generally, this trend is confirmed.

A discussion of the systematic errors in the large-aperture analysis
was  given in Paper I.  Many potential problems were considered, e.g.,
poor scaling of continuum images to line$+$continuum images or poor sky
subtraction.  In general, such errors result in  shifts of a few 0.1
Myr in the entire IB age scale, with little effect on the sequence of
ages in the galaxy.

A few details of the flux determinations and the models are as follows:

\begin{enumerate}
\item  No background flux internal to NGC 4449 is subtracted, since this
procedure was shown in Paper I to have little effect on the IB ages.

\item  Extinction is computed for each bandpass using the algorithm of
\cite{calz4}, described above (\S \ref{sec:obs:ebv}).

\item  For simplicity, the IB models are computed with an initial mass
function (IMF) over the range $0.1-110$ M$_{\odot }$ with a Salpeter
slope, whereas a multi-component IMF was used in Paper I.
\end{enumerate}

The photometric results are given in Table \ref{tab:lgap}, in which the
columns are as follows: (1) ordinal number or object name; (2)
location: {\em N} for the northern part of the galaxy, {\em B} for the
bar, and {\em O} for other areas; (3) logarithm of the spectral
irradiance (flux) in the UIT B1 band in ergs cm$^{-2}$ s$^{-1}$ \AA
$^{-1}$; (4) logarithm of the irradiance in the H$\beta $ line in ergs
cm$^{-2}$ s$^{-1}$; (5) H$\beta $ equivalent width in \AA~ (no
background within NGC 4449 subtracted); (6) logarithm of the irradiance
in the H$\alpha$ line in ergs cm$^{-2}$ s$^{-1}$; (7) H$\alpha$
equivalent width in \AA~ (no background within NGC 4449 subtracted);
(8) color index computed from the UIT B1 band and the narrowband
continuum at 4830 \AA, respectively, in magnitudes (same $f_\lambda$
zero point for both bands); (9) color index computed from the
narrowband continuum at 4830 \AA~ and 6520 \AA, respectively, in
magnitudes (same $f_\lambda$ zero point for both bands); (10) age in
Myr computed from $\log N_{{\rm Lyc}}/L_{{\rm B1}}$ using the IB
cluster model.

Internal errors in FUV fluxes are $\sim 1-3\%$; those in optical fluxes
are mostly less than $1\%$, except for a few H$\beta$ measurements with
errors of $5-7\%$.  The internal errors in the age, computed by
propagating the photometric errors through extinction correction and
the IB model, are less than 0.1 Myr, except for sources 1, 3, 7, 15,
19, 20, 21, and 22, which have errors of $0.1 - 0.4$ Myr.

The IB ages inferred from the ratio H$\alpha$/B1 are depicted in a bar
chart in Figure \ref{fig:bar}. Again, the youngest OB association
complexes in terms of IB models are in a group located along the
northeastern edge of the galaxy. For reference, a 2.5 Myr difference in
IB age requires H$\alpha$/B1 ratios to differ by a
factor of 3.

The observed B1/blue flux ratio for the large apertures is plotted vs.
H$\alpha$ emission equivalent width, denoted EW(H$\alpha$), in Figure
\ref{fig:ewha}, where the apertures are classified by their location in
the galaxy into the categories {\em north} (triangles), {\em bar}
(squares), and {\em other} (plusses).  Both the plotted quantities
decrease with age.  All but one of the northern apertures lie to the
upper right of the plot, with log B1/blue$>0.7$ and
EW(H$\alpha$)$>110$\AA, i.e., where both H$\alpha$ and FUV are strong
compared to the optical continuum.

The H$\alpha$/FUV ratio is interpreted in this paper as being
determined by star formation history.  However, two possible objections
to this idea are the following:  (1) differences in H$\alpha$/FUV can
be caused by differences in the amount of extinction, since a
relatively high extinction suppresses FUV more than H$\alpha$;  (2) 
differences in H$\alpha$/FUV can be caused by differences in the
leakage of the ionizing photons from the H{\sc ii} regions. Figure
\ref{fig:ewha} weighs against these objections.  If the high values of
H$\alpha$/FUV in the north are caused by extinction, the same sources
should also have low FUV/blue ratios, which is not the case.
Furthermore, EW(H$\alpha$) is a ratio of two quantities at nearly the
same wavelength and thus is unaffected by reddening.  One could
postulate a difference in dust geometry between the sources in the
north and the others.  E.g., low values of the parameter
$E(B-V)_s/E(B-V)_g$ could explain the high B1/blue ratios in the north,
since both B1 and blue are stellar continuum; however, this would lower
EW(H$\alpha$) by raising the stellar continuum in relation to
H$\alpha$.  Finally, although variations in ionizing flux leakage could
explain variations in EW(H$\alpha$), the B1/blue ratio is immune to
this effect.

In other words, we find a fourfold correlation between position in the
galaxy, H$\alpha$/B1, B1/blue, and EW(H$\alpha$), which is 
evidence for a real difference between regions.  Certainly, it is
possible to find complex hypotheses that would explain these 
correlations without invoking age or star formation history.  E.g., the
extinction curve in the north could be flatter than in the bar, thus
causing B1/blue to be higher in the north, and this effect could be
combined with a large difference in ionizing flux leakage. 
Nevertheless, such a scenario is significantly more subtle than the two
simple objections noted above.  Surely, the either/or interpretation at
the end of \S\ref{sec:sfr:lgap:rec} offers the best working hypothesis.

If, indeed, the difference in IB model age between the north and the
bar is mainly caused by a difference in the {\em type} of
star-formation history, i.e., IB vs. CSF, then this interpretation is
compatible with that offered by \cite{huntsf} that star formation has
occurred over a long period of time in the periphery of NGC 4449, just
as it has in the bar, but with relatively infrequent and distinct
episodes, of which the present episode is one instance.

Using various photometric bands, the average star formation rate (SFR)
over a timescale appropriate for each band is computed for the whole
galaxy from CSF models.  This timescale is given by a flux-weighted mean
of the ages of IB models, as follows:
\begin{displaymath} <t>_\lambda = \frac{\int_0^\infty t
L_{\mathrm{IB}}(\lambda,t) dt} {\int_0^\infty
L_{\mathrm{IB}}(\lambda,t) dt} {\rm Myr}, 
\end{displaymath}
where $L_{\mathrm{IB}}(\lambda,t)$ is the luminosity at a given
wavelength computed using the IB model for age $t$ Myr. The results
are given in Table \ref{tab:sfr} (cf. Table 6 of Paper I). Each row is
for one bandpass and its associated timescale, given in the first two
columns.  Each of the three columns labeled ZERO, CALZ, and GAL,
respectively, gives the results of a different method of extinction
correction.  The SFR values under ZERO have no correction for extinction
within NGC 4449,
those under CALZ are corrected using the \cite{calz4} method for the
NGC 4449 component as described in \S\ref{sec:obs:ebv}, and those under
GAL are corrected by using the Galactic curve for both the NGC 4449 and
foreground components.

Table \ref{tab:sfr} shows that the SFR over last few $10^{7}$ yr (UIT
B1 band) is the most sensitive to the extinction correction of the
three timescales considered.  Both the ZERO and the CALZ results are
fairly
consistent over the three timescales; however, the extinction in NGC
4449 is substantial, and a correction is needed, so the SFR values
under ZERO are rejected.  The GAL method is simple, but it produces an
apparent $\sim 5-$fold peak in the SFR  $\sim 10^{7}$ yr ago, because
of the strong enhancement in the derived FUV fluxes with respect to
H$\alpha$ and blue continuum.  This alone is not sufficient reason to
reject the method, since such an enhancement might occur.  
Nevertheless, we provisionally adopt the scenario of consistent SFRs as
the more likely one and for that reason, we use the \cite{calz4}
extinction-correction method.  In Paper I, the same goal was achieved
by an {\em ad hoc} use of the Orion Nebula extinction curve, which has
a shallow FUV rise (\cite{orion}).

The extinction-corrected H$\alpha$ photometry can be used to compare
sources in NGC 4449 to the local standard for bright H{\sc ii} regions,
30 Dor.  The value of $L_{{\rm H}\alpha}$ adopted for 30 Dor is
$10^{40.2}$ ergs s$^{-1}$ (\cite{conti}).  With an extinction of
$E(B-V)=0.43$ and a distance of 5.4 Mpc, this value becomes
$10^{-11.76}$ ergs s$^{-1}$ cm$^{-2}$, where we use the Galactic
extinction curve.  Comparison with Table \ref{tab:lgap} shows that only
one large aperture includes a  greater H$\alpha$ luminosity than 30
Dor, i.e., aperture 18, which encloses a complex region in the southern
part of the bar, containing several clusters and H$\alpha$ shells.  Let
$n_{{\rm 30D,H\alpha}}$ be the number of 30 Dor units of H$\alpha$ flux
emitted by a given source. Then, for aperture 18, $n_{{\rm
30D,H\alpha}}=1.9$; for large aperture 6, which includes the brightest
compact H{\sc ii} region in NGC 4449, $n_{{\rm 30D,H\alpha}}=0.8$; for
the whole of NGC 4449, $n_{{\rm 30D,H\alpha}}=11.5$.  For the LMC
itself, if we adopt an H$\alpha$ luminosity of $4.1\times10^{40}$ ergs
s$^{-1}$ (\cite{kennlmc}), then $n_{{\rm 30D,H\alpha}}\approx 2.5$.

\subsection{Small Apertures\label{sec:sfr:smap}}

The UIT B5 image has spatial resolution of $\sim 3\arcsec$, equivalent
to a projected distance of $\sim 80$ pc in NGC 4449.  In the LMC, this
size is of the same order as the diameter of a small OB association
(\cite{lh}).  Accordingly, compact knots consisting of a few resolution
elements in the B5 image of NGC 4449 can be treated reasonably as
coeval OB associations.  Although this assumption is unlikely to be
strictly accurate (\cite{assoc}), it is both commonly applied and
difficult to avoid in OB association studies.

We compute ages for our small associations from H$\alpha$/B5 flux
ratios in circular apertures.  The definition of apertures is done with
a program that displays the B5 and the H$\alpha$ images side by side. A
circular aperture is overplotted on each image and sized interactively.
The purpose of the simultaneous sizing is to try to define a single
aperture containing associated stars and gas in both bands.  A
background annulus around each aperture is also defined simultaneously
for both images in the same way.  The resulting apertures are
overplotted on the B5 image in Figure \ref{fig:smb5} and on the
H$\alpha$ image in Figure \ref{fig:smha}.  The numbers are shown in
Figure \ref{fig:smap}, and the locations and sizes are given in Table
\ref{tab:smpos}.

The goal in defining the small apertures is to characterize the
compact, bright FUV sources seen within larger structures, such as the
bar, the northern ``arms,'' and the southern extensions.  We include
mainly sources that are bright and in good contrast against the
background in both bands; but some sources that are bright in only one
band are included if they are in the immediate neighborhood of sources
seen well in both bands.  E.g., sources 17, 18, and 20 are conspicous
in the B5 image and faint in H$\alpha$, whereas the reverse is true of
source 41, yet all of these sources are included because they are part
of larger structures that include several other measurements.  However,
a number of compact H{\sc ii} regions are both relatively isolated and
lacking a distinct counterpart in the B5 image, and so they are
omitted.

Compact FUV sources in several external galaxies have been analyzed in
WF/PC imagery by \cite{meurer}, who investigate the distribution of
star formation in central starburst galaxies and conclude that super
star cluster formation is $\sim 20\%$ a of the total SFR. 
Unfortunately, the UIT resolution does not permit a direct comparison
of our results to those achieved with WF/PC.  The sources of
\cite{meurer} have diameters of order $\sim 10$ pc and in many cases
less, in contrast to the 80 pc limit imposed by the resolution of our
B5 image.  Moreover, we infer typical 2200 \AA~ absolute magnitudes
$-16 \gtrsim M_{220} \gtrsim -18$ for our small apertures, based on the
color B5$-220 \approx -0.9$ for a 4 Myr old IB model.  These values are
$1-2$ mag brighter than most of those in the WF/PC study, showing again
that our small sources are more likely to be associations than
clusters.  However, the real size and morphology of these sources
cannot be determined from the UIT data.

Background subtraction within NGC 4449 is a difficult problem because
of crowding and surface brightness gradients.  Accordingly, we do not
apply  the background annuli individually to the particular sources
they surround, but instead, we use them in a more complex procedure.
The sources are divided into groups, each of which appears likely to
have a common background value.  Each background annulus is broken up
into octants, and all the octants for a group of sources are treated
together, regardless of which source each octant was originally
associated with.

The photometry program integrates the fluxes for each aperture and for
each background octant.  Each of these fluxes is separately corrected
using its own Balmer decrement with the method described
above (\S\ref{sec:obs:ebv}).  The collection of octants for each group
of sources is sorted, and the $3/8$ of the collection with the lowest
values are selected.   The mean of these lowest values is adopted as
the background for the group of sources, and the standard error of the
mean is adopted as the error in the background.  

The photometric and modeling results are given in Table \ref{tab:smap},
in which the columns are as follows: (1) ordinal number of source; (2)
group membership, with each group assumed to have a single, common
value of background flux;  (3) logarithm of the spectral irradiance in
the UIT B5 band in ergs cm$^{-2}$ s$ ^{-1}$ \AA $^{-1}$; (4) logarithm
of the irradiance in the H$\beta $ line in ergs cm$^{-2}$ s$^{-1}$; (5)
logarithm of the irradiance in the H$\alpha$ line in ergs cm$^{-2}$
s$^{-1}$; (6) $E(B-V)_g$ in magnitudes, including Galactic foreground;
(7) logarithm of the inferred ionizing photon luminosity in photons
s$^{-1}$; (8) logarithm of the inferred spectral luminosity in the B5
band in ergs s$^{-1}$ \AA $^{-1}$; (9) IB age in Myr as computed from
the H$\alpha$/B5 ratio (the value $\mathit{<0}$  means that $\log
N_{{\rm Lyc}}/L_{{\rm B5}}$ is greater than the range covered by the
models); (10) estimated error in age in Myr; (11) note, encoded as
follows:  {\em Bn}, H$\beta$ background less than zero; {\em Un}, FUV
background less than zero; {\em Bx,} source flux less than zero in
H$\beta$.

As a check of the photometry, we have looked for a correlation between
the $E(B-V)$ values of all the sources and their IB ages, to see
whether the age variation merely reflects errors in the reddening
estimates, because of the leverage of the FUV rise in the extinction
correction, but no such correlation is found.

Background fluxes less than zero are subtracted from the flux of the
source just like ones greater than zero, so that a minus background
raises the flux of the source; such cases are assumed to have resulted
from a correction for continuum and overall sky that is too great in
the particular region.  However, source fluxes less than zero are
omitted from subsequent analysis.

Internal errors in the raw small aperture fluxes are typically $\sim
6\%$ for B5, $\sim 1\%$ for H$\alpha$, and $\sim 3\%$ for H$\beta$. The
maximum errors found for individual sources are $20\%$ for B5, $5\%$
for H$\alpha$, and $13\%$ for H$\beta$.  Error in the background value
for each set of grouped small apertures contributes significantly to
the total photometric error.  The resulting error in the age estimate
is computed from the following
propagation-of-error formula:   
\begin{displaymath} \sigma^2({{\rm
age}})  = 7.49 (\sigma(F_{{\rm H\alpha}})/F_{{\rm H\alpha}})^2  +
31.39(\sigma(F_{{\rm H\beta}})/F_{{\rm H\beta}})^2  + 8.17
(\sigma(f_{{\rm B5}})/f_{{\rm B5}})^2.   
\end{displaymath}  
This formula takes into account the adopted extinction correction
method and the dependence of the computed extinction on H$\alpha$ and
H$\beta$, and it uses an approximate linear fit to the model trace
shown in Figure \ref{fig:model}.  The H$\alpha$ contributes less to the
total error than either of the other measurements, because it enters
the computation twice, in ways that cancel.  With other values held
constant, an increase in H$\alpha$ leads to an increase in age via the
extinction correction, which raises the FUV flux, but the same increase
in H$\alpha$ leads to a decrease in age directly via the IB model.

The correspondance between small and large apertures is given in Table
\ref{tab:lgsm}, together with the IB age and stellar reddening
$E(B-V)_s$ (\S \ref{sec:obs:ebv}) for each small aperture.  In most
cases, the large-aperture age is at least as great as the youngest
small-aperture age in the corresponding group, as expected if the
compact sources contain the youngest stars.  However, for large
apertures 1, 7, 15, 18, 21, and 22, this relationship does not hold,
i.e., the large-aperture age is less than at least one corresponding
small-aperture age by 1 Myr or more.  For large apertures 7 and 18, the
reason may be crowding of compact sources.  For large aperture 15, the
individual compact sources are aging OB assocations and are probably a
minor source of the ionizing flux that causes the surrounding
H$\alpha$ emission.  Large apertures 1, 21, and 22 contain diffuse
filamentary H$\alpha$ features, and thus account better for the
ionizing flux than do the small apertures.

A related analysis of stellar generations  uses FUV spectra taken with
the $10\arcsec \times 20\arcsec$ aperture of {\em IUE} (\cite{athome};
\cite{thesis}).  The pointings correspond to our large apertures 6, 10,
14, 16, 18, 19, and 21.  The analysis is an evolutionary synthesis
different from the one done in this paper, in that it is based on a
grid of binned high-resolution {\em IUE} stellar spectra.  The fitting
depends primarily on age-sensitive spectral features, and allows either
one or two age components.   These results show evidence for multiple
generations in the age range $0-12$ Myr at 5 of the 7 positions.  The
youngest mean age is given by large aperture 6, consistent with our
results.  Typically, the {\em IUE} ages are $\sim 3$ Myr greater than
our corresponding large-aperture ages.  Reddenings $E(B-V)$ within NGC
4449 of a few $0.01$ are found by fitting the {\em IUE} spectra of the
OB associations with moderately steep, LMC or 30 Dor extinction curves
(\cite{thesis}).  These reddenings are lower than the values $E(B-V)_s
= 0.44 E(B-V)_g$ in our small apertures at the same positions by
$\Delta E(B-V) \approx 0.1$.  Hence, the {\em IUE} analysis is
consistent with $E(B-V)_s < E(B-V)_g$.  Taken at face value, our
results imply that the proportionality constant between $E(B-V)_s$ and
$E(B-V)_g$ is less than 0.44; however, the apertures are not matched
between the two sets of data.

Figure \ref{fig:smap} shows a map of our compact source ages, with
outlines emphasizing 3 regions that show a spatial progression of ages.
These patterns may indicate sequential star formation (SSF;
\cite{elmelada}; \cite{thesis}).  In this scenario, the typical spatial
separation between stellar generations appears to be $\sim 100-300$ pc,
a factor of $5$ larger than in Elmegreen \& Lada's Galactic
examples.  In the FUV-brightest region of the southern bar (1.0, -0.5
in Figure \ref{fig:smap}, corresponding to large aperture 18 in Figure
\ref{fig:lgap}), we measure 6 sources, with a progression of ages from
the west (young, sources 41 and 43) to the east (old, source 45).
Sources 43 and 45 lie approximately along an east-west line with a
separation of $0\farcm13$, or $204$ pc.  The age difference of 8.3 Myr
implies a propagation speed of $25$ km s$^{-1}$, which is larger than
the canonical sound speed of 10 km s$^{-1}$ for 8000 K H{\sc ii}
regions (\cite{spitzer}); however, this speed is less than that of an
associated ionized shell (40 km s$^{-1}$), and in the same range as
other shells in NGC 4449 (\cite{hg90}).  This group is physically the
most likely case of SSF, as the distances between generations are
relatively small, and the propagation velocity is moderate.  Moreover,
a double H$\alpha$ shell emanating from this group toward the east, at
coordinates $(0.9, -0.6)$ in Figure \ref{fig:smha}, is suggestive of
multiple, discrete star formation episodes. 

In the northwest ``arm'' of NGC 4449, a group of 9 sources shows a
progression of ages from west (young, sources 12 and 14) to east (old,
sources 17 and 22).  Small apertures 14 and 17, with an age difference
of 10.4 Myr, are separated by 756 pc, implying a propagation speed of
73 km s$^{-1}$.   A different pattern is seen in the center of the
galaxy, where a line of young associations (33, 34, 38, 39, 47, 
49, and 50), parallel to the bar,  lies $\sim 0\farcm3$ east of the older
associations along the ridgeline (31, 32, 36, 37, 45).  The propagation
speed between these two lines would be $\sim 70-100$ km s$^{-1}$.

Models of propagating star formation in NGC 4449 are computed by
\cite{thesis}.  The models are of two types: ``short-range,'' in which
propagation is by the ionization-shock front of an expanding H{\sc ii}
region, and ``long-range,'' in which propagation is by wind- and
supernova-blown bubbles.  The long-range case is more relevant to our
observations.  Here, the propagation is modeled in two phases: a bubble
is pressurized by winds and supernovae for 3 Myr, then begins an
adiabatic free expansion, which continues until the surrounding shell
fragments.  The expansion velocity of the bubble after the first phase
is an upper limit on the total propagation speed.  The outcome of the
models is that this limit is in the range $20-50$ km s$^{-1}$ for the 7
{\em IUE} pointings described above, comparable to that of shell
velocities observed in NGC 4449 (\cite{hg90}).

A few other instances of possible SSF outside our Galaxy are found in
the literature.  One recent example is from an infrared study of the
center of M82 (\cite{m82}), where a linear arrangement of clusters with
approximately ordered ages extends over a distance of $\sim 230$ pc.
The implied propagation speed is $\sim50$ km s$^{-1}$, consistent with
the gas motions in that environment.  A smaller-scale example is the OB
associations LH 9 and LH 10 in the LMC (\cite{lh9}), which are $\sim
5\arcmin$ apart, or $\sim 70$ pc for a distance modulus of 18.48
(\cite{wester}).  Although the authors do not commit themselves to a
definite value for the age difference between the two associations, from
the upper mass limits of their color-magnitude diagrams, it appears to
be small, $\sim 1$ Myr or less, using the Geneva stellar evolution
tracks (\cite{tracks}).  The implied propagation speed is $\sim 70$ km
s$^{-1}$.

In summary, in some regions of NGC 4449, the OB associations appear to
be arranged spatially in an approximate order of age.  In two of the
three cases discussed above, the implied propagation speeds are
somewhat higher than typical measured H$\alpha$ shell velocities in NGC
4449, though within a factor of $2$ of apparent propagation speeds
seen in other galaxies.  On the other hand, the region of large
aperture 18 contains the most likely instance of propagating star
formation, as the velocity of propagation would be in the range
expected for the motion of expanding ionized shells.

\section{Conclusions\label{sec:conc}}

Color maps in H$\alpha$/FUV and FUV/blue continuum show variation in
these ratios over the galaxy.  The northern tier of H{\sc ii} regions
shows a generally high H$\alpha$/FUV ratio compared to the rest of the
galaxy.  The regions of high FUV/blue ratio coincide with outlying OB
complexes, but the OB complexes in the nuclear region, which has a high
optical surface brightness, are in low contrast against the
background in the FUV/blue map.

The scenario previously presented in Paper I, i.e., that the youngest
UV-emitting populations are found in the northern part of NGC 4449, is
reinforced.  However, young associations also exist in the bar region
of the galaxy.   Photometry of large OB complexes is consistent with
the idea of \cite{huntsf} that star formation occurs continually in the
periphery of NGC 4449, just as it does in the bar, but with relatively
infrequent and distinct episodes.

Photometry of compact sources in small apertures shows several
instances of correlation between age and spatial location, but the
physical significance of these correlations is unclear.  Direct
propagation of star formation from one site to the next appears
possible in at least one case.  The implied propagation speeds are in
the range $20-100$ km s$^{-1}$. 

The interpretation of FUV aperture photometry in NGC 4449 relies on 
the starburst extinction model of \cite{calz4}.  If the consistency of
average star formation rates over several time intervals is used as
a constraint, then this treatment of extinction appears justified for
NGC 4449.

The radial surface brightness profile in the FUV approximates an
exponential disk, in contrast to the similar galaxy NGC 4214, where the
FUV profile follows an $r^{1/4}$ law.  The ratio of diffuse FUV flux to
H$\alpha$ flux in the smooth, extended component of NGC 4449 is
consistent with photoionization.

\acknowledgements 

Thanks to Bill Waller for commenting on the manuscript.  Thanks to
Allen Home for a copy of his informative thesis.  The authors are
grateful to all those on board the Shuttle for both Astro-1 and
Astro-2, as well as to the engineers and technicians who made these
missions possible.  Funding for the UIT project has been through the
Spacelab Office at NASA Headquarters under project number 440-51.  R.
W. O. was supported in part by NASA grants NAG 5-700 and NAGW-2596 to
the University of Virginia.

\label{bibstart}

\clearpage

\figcaption{Ultraviolet and ground-based images of NGC 4449 (see Table
\ref{tab:obs}).  North is up and east to the left, with coordinates in
arcmin. Power-law gray scaling with index 0.3.  UIT images in broad FUV
bands (B1 and B5) are on the left, and GAFPIC images in narrow optical
bands are on the right.  B1 and B5 have mean wavelengths of 1521\AA\
and 1615\AA, respectively, for a flat input spectrum.  {\em Blue} is
the off-band continuum at 4830\AA, near H$\beta$.  H$\alpha$ is
continuum-subtracted.  Four foreground stars are seen in the optical
data at $(-1.6, -0.8)$, $(-1.6, 1.4)$,  $(-0.8, -1.7)$, and $(0,0)$;
these are imperfectly subtracted in the H$\alpha$ image because of a
slight PSF mismatch. \label{fig:all}}

\figcaption{Contour map of UIT FUV Image of NGC 4449 in the B1 Band. 
North is up and east to the left, with coordinates in arcmin.  In units
of $10^{-17}$ ergs cm$^{-2}$ s$^{-1}$ \AA$^{-1}$ arcsec$^{-2}$, the
contour levels are 1.3, 3.9, 11.6, 34.8, and 104.5.   \label{fig:cont}}

\figcaption{IB cluster models used to derive ages of UV and H$\alpha$
sources.   A Salpeter IMF is assumed with mass limits $0.1 M_\odot$ and
$110 M_\odot$.  Abscissa: time since the formation of a coeval group of
stars.  The left-hand ordinate axis gives the log of the ratio of the
ionizing photon luminosity (photons s$^{-1}$) to the FUV luminosity
(ergs s$^{-1}$). The right-hand ordinate axis gives the log of the
observable ratio H$\alpha$/B1 assuming our average extinction model (\S
\ref{sec:obs:ebv}). \label{fig:model}}

\figcaption{Map of log H$\alpha$/B1.  North is up and east to the
left, with coordinates in arcmin.  Flux ratio scaling from log
H$\alpha$/B1 of 0.2 (white; less H$\alpha$) to 1.6 (black; more
H$\alpha$); the actual range of values in the galaxy is $0.25-1.98$. 
H$\alpha$ is smoothed by a Gaussian of FWHM $3\farcs 6$.  To remove sky
noise, H$\alpha$ values are clipped at $1.6\times10^{-16}$ ergs
cm$^{-2}$ s$^{-1}$ arcsec$^{-2}$, and B1 at $2.2\times10^{-17}$ ergs
cm$^{-2}$ s$^{-1}$ \AA$^{-1}$ arcsec$^{-2}$.  See Figure \ref{fig:all}
for coordinates of foreground stars. \label{fig:rathau}}

\figcaption{IB cluster evolution models, aged from 1 Myr to 1 Gyr.  A
Salpeter IMF is assumed with mass limits $0.1 M_\odot$ and $110
M_\odot$.  Dashed line: log of ratio of H$\alpha$ flux to FUV flux in
the UIT B1 band; solid line: log of ratio of B1 to blue continuum. 
Right-hand ordinate axes give observable ratios H$\alpha$/B1 and
B1/blue assuming our average  extinction model (\S \ref{sec:obs:ebv}). 
\label{fig:long}}

\figcaption{Map of log B1/blue.  North is up and east to the left, with
coordinates in arcmin.  The flux ratio scaling is from log B1/blue of
$0.04$ (light; less UV) to 1.5 (dark; more UV).  The actual range of
values in NGC 4449 is $-0.29$ to $+1.25$.  Light circles are foreground
stars.  The blue (4830 \AA) is smoothed with a Gaussian of FWHM
$3\farcs 6$, then both images are additionally smoothed with a Gaussian
of FWHM $2\farcs3$.  To remove sky noise, B1 values are clipped at
$2.6\times10^{-17}$ ergs cm$^{-2}$ s$^{-1}$ \AA$^{-1}$ arcsec$^{-2}$,
and blue at $9.6\times10^{-18}$ ergs cm$^{-2}$ s$^{-1}$ \AA$^{-1}$
arcsec$^{-2}$.  \label{fig:ratuvb}}

\figcaption{Surface brightness profiles of NGC 4449 in the UIT B1 band
and in the 6520 \AA~ narrow red continuum band.  Abscissa:  mean radius of
annular aperture in arcsec, or 0.5 the radius of the central circular
aperture.  Ordinate:  log of the surface brightness in units of ergs
cm$^{-2}$ s$^{-1}$ \AA$^{-1}$ arcsec$^{-2}$.  Outside the central cusp,
the $r$ band fits an exponential profile except for a hump at a radius
of $\sim 100^{\prime \prime }$, due to the bright star-forming complex
at the northern edge of the galaxy. Photometric errors are given by the
error bars for B1 and are less than the size of the symbols in
$r$.\label{fig:sbprof}}

\figcaption{Froth regions marked on continuum-subtracted GAFPIC
H$\alpha$ map; photometry is given in Table \ref{tab:froth}.  North is
up and east to the left, with coordinates in arcmin.  See Figure
\ref{fig:all} for coordinates of foreground stars, which are
imperfectly subtracted because of a slight PSF mismatch. 
\label{fig:froth}}

\figcaption{UIT FUV Image of NGC 4449 in the B5 band. North is up and
east to the left, with coordinates in arcmin.  Logarithmically scaled.
Large apertures from Paper I are indicated (Table \ref{tab:lgpos}).
Sources are classified as follows according to the region of the galaxy
in which they are found. North: 1, 2, 4, 5, 6, 7, 9, 13, 15, 17, 19;
bar: 8, 10, 11, 12, 14, 16, 18; other: 3, 20, 21, 22. \label{fig:lgap}}

\figcaption{Bar chart showing IB ages computed from B1 and H$\alpha$
fluxes in large apertures (Table \ref{tab:lgap}). Abscissa: large
aperture number. Ordinate: age in Myr. Region of the galaxy where the
sources are located is encoded in shading of the bars as given by
legend (cf. Figure \ref{fig:lgap}). The youngest 6 sources are along
the northeastern edge of the galaxy. \label{fig:bar}}

\figcaption{Plot of log B1/blue vs. equivalent width in H$\alpha$ for
large aperture measurements with no extinction corrections.  Plot
symbols signify region of the galaxy as in Table \ref{tab:lgap} and
Figure \ref{fig:lgap}: triangles, north; squares, bar; plusses, other.
\label{fig:ewha}}

\figcaption{Small apertures on the UIT B5 image.  North is up and east
to the left, with coordinates in arcmin.  Apertures are defined
manually by adjusting them simultaneously on B5 and H$\alpha$ images. 
Cf. Figure \ref{fig:smha}.  Aperture positions and sizes are given in
Table \ref{tab:smpos}; numbering and modeled ages are given in Figure
\ref{fig:smap}.  Image is log scaled, and pixel values are clipped to be
greater than a surface brightness of $1.6\times10^{-17}$ ergs cm$^{-2}$
s$^{-1}$ \AA$^{-1}$ arcsec$^{-2}$.  \label{fig:smb5}}

\figcaption{Small apertures on the H$\alpha$ image. North is up and
east to the left, with coordinates in arcmin.  Apertures are defined
manually by adjusting them simultaneously on B5 and H$\alpha$ images. 
Cf. Figure \ref{fig:smb5}.  Aperture positions and sizes are given in
Table \ref{tab:smpos}; numbering and modeled ages are given in Figure
\ref{fig:smap}. Image is log scaled, and pixel values are clipped to be
greater than a surface brightness of $1.6\times10^{-16}$ ergs cm$^{-2}$
s$^{-1}$ arcsec$^{-2}$. \label{fig:smha}}

\figcaption{Small aperture ages on the UIT B5 image.  North is up and
east to the left, with coordinates in arcmin.  Numbers are as in Tables
\ref{tab:smpos} and \ref{tab:smap}.  Symbols denote ages in Myr, as
follows: $0-3$, triangles; $3-5$, squares; $5-7$, plusses; $>7$,
crosses. Source 55, which does not have a measurable H$\beta $ flux, is
included in the oldest category.  Outlines emphasize  regions where
ages are in approximate sequence. E.g., ages decrease from east to west
starting with source 45. \label{fig:smap}}

\newpage

\begin{table}[tbp] \centering
\caption{Observations}\label{tab:obs} 
\end{table}

\begin{table}[tbp] \centering
\caption{Comparison of Absolute Magnitudes}\label{tab:integ} 
\end{table}

\begin{table}[tbp] \centering
\caption{Photometry of Froth Regions}\label{tab:froth} 
\end{table}

\begin{table}[tbp] \centering
\caption{NGC 4449 Large Apertures}\label{tab:lgpos} 
\end{table}

\begin{table}[tbp] \centering
\caption{Photometry in Large Apertures}\label{tab:lgap} 
\end{table}

\begin{table}[tbp] \centering
\caption{Global Star Formation Rates}\label{tab:sfr} 
\end{table}

\begin{table}[tbp] \centering
\caption{NGC 4449 Small Apertures}\label{tab:smpos} 
\end{table}

\begin{table}[tbp] \centering
\caption{Photometry in Small Apertures}\label{tab:smap} 
\end{table}

\begin{table}[tbp] \centering
\caption{Correspondance of Large and Small Apertures}\label{tab:lgsm} 
\end{table}

\end{document}